\documentclass[prl,aps,showpacs,twocolumn,floatfix]{revtex4}
\usepackage{graphics,bm}
\usepackage{amssymb}
\usepackage{epsfig}
\usepackage{epsf}
\usepackage[usenames]{color}
    
\usepackage{graphicx}% Include figure files
\usepackage{dcolumn}% Align table columns on decimal point
\usepackage{bm}% bold math

\begin{document}

\preprint{APS/123-QED}

\title{What Determines the Wave Function of Electron-Hole Pairs in Polariton Condensates?}
% Force line breaks with \\
\author{Kenji Kamide}
\email{kamide@acty.phys.sci.osaka-u.ac.jp}
\author{Tetsuo Ogawa}%
\affiliation{%
Department of Physics, Osaka University, Toyonaka, Osaka 560-0043, Japan
}%
%\affiliation{

\date{\today}% It is always \today, today,

\begin{abstract}
The ground state of a microcavity polariton Bose-Einstein condensate is determined by considering experimentally tunable parameters such as excitation density, detuning, and ultraviolet cutoff.
During a change in the ground state of Bose-Einstein condensate from excitonic to photonic, which occurs as increasing the excitation density, the origin of the binding force of electron-hole pairs changes from Coulomb to photon-mediated interactions.
The change in the origin gives rise to the strongly bound pairs with a small radius, like Frenkel excitons, in the photonic regime.
The change in the ground state can be a crossover or a first-order transition, depending on the above-mentionsed parameters, and is outlined by a phase diagram.
Our result provides valuable information that can be used to build theoretical models for each regime.
\end{abstract}

\maketitle

Microcavity polaritons---photoexcited electrons and holes strongly coupled with photons in a semiconductor microcavity---have been observed to exhibit Bose-Einstein condensation (BEC)~\cite{Kasprzak1,Balili}.
Due to the light-matter coupling, the polariton has an extremely small mass about $10^{-4}$ times the free-electron mass; the small mass results in a high critical temperature and low critical density.
BEC can be realized even at room temperature~\cite{Baumberg}, which is remarkable considering that it had been difficult to realize BEC in exciton systems for a long time~\cite{Moskalenko}.
Microcavity polaritons are dissipative particles due to the short lifetime of photons and inelastic scattering of excitons by phonons. Therefore, the polariton BEC is in a nonequilibrium stationary state with a balance between pumping and losses~\cite{Carusotto,Szymanska,Wouters}. However, the polariton BEC has many similarities with BEC of neutral atoms in a thermal equilibrium~\cite{Pitaevskii}.
It shows several evidences for the superfluidity: the Goldstone mode~\cite{Utsunomiya}, the quantized vortices~\cite{Lagoudakis}, and the collective fluid dynamics~\cite{Amo}.

Such a stationary state appears to be well described by the ground state of a closed microcavity polariton system, when the polariton lifetime is longer than the thermalization time~\cite{Szymanska,Kasprzak2}. In this paper, assuming such a situation, we determine the ground state with a fixed excitation density at absolute zero as a function of experimentally variable parameters: excitation density, detuning~\cite{Bajoni}, and ultraviolet cutoff determined by the lattice constant.
In past studies, mean-field theories have been used to discuss the two limits---low excitation density~\cite{Eastham,Keeling} and high excitation density~\cite{Marchetti1}---by considering two different models. These theories are complementary~\cite{Marchetti2}, but their relation is somewhat ambiguous.
We investigate the intermediate density region as well, where the electron-hole (eh) wave function of the relative motion becomes important.
We show that the ground state energy and wavefuction gradually change from those of excitons to photons as the excitation density increases. It is also shown that the change can be a crossover or a first-order transition, depending on the parameters considered.
The latter transition is characterized by a jump in the photonic fraction, and a sudden narrowing of the eh wave function that is accompanied by a change in the binding force from Coulombic force to photon-mediated force.

The polariton system can be described as electrons and photons with a total excitation number $N_{\rm ex}$, which are interacting through electric dipole coupling. The Hamiltonian is given by $H=H_{\rm el}+H_{\rm el-el} +H_{\rm ph}+H_{\rm el-ph}$~\cite{Marchetti1},
\begin{eqnarray}
H_{\rm el}&=&\sum_{k}\varepsilon_{e}(k) a^\dagger_k a_k + \varepsilon_{h}(k) b_k b^\dagger_k, \\ 
H_{\rm el-el}
%= \sum_q \frac{1}{2}V_q
%\left( \rho_q \rho_{-q}-N_{\rm e}-N_{\rm h} \right)
&=&\sum_{q}  U_q : \rho_q \rho_{-q} :,\\
H_{\rm ph}&=&\sum_{k}\left( \sqrt{(ck)^2+(\hbar \omega_{\rm c})^2} -\mu \right) \psi_k^\dagger\psi_k, \\
H_{\rm el-ph}&=& -g \sum_{k,q} (\psi_q a^\dagger_{k+q}b_k +\psi^\dagger_q b^\dagger_{k+q}a_k ), \label{eq:dipole}
\end{eqnarray}
where $\varepsilon_{e,h}(k)\left[ =\hbar^2 k^2/2m_{e,h}+(E_{\rm g}-\mu)/2 \right]$ and $U_q \left[ =(2 \pi e^2/\epsilon^\ast V q^2) \right]$ are the electronic dispersion in an effective mass approximation and the Coulomb interaction, respectively.
Further, $a_k$, $b_k$, and $\psi_k$ are the annihilation operators of the conduction and valence electrons and photons with momentum $k$, respectively. Fourier transform of the density operator given by $\rho_q =\sum_k (a^\dagger_{k+q} a_k -b_{k-q} b^\dagger_k )$.
The zero-point frequency of the cavity photons is $\omega_{\rm c}$ and detuning is defined as $d=(\hbar \omega_{\rm c}-E_g)/\varepsilon_0$, where $\varepsilon_0$ is the exciton Rydberg. The light-matter coupling constant is given by $g = d_{\rm cv} \sqrt{2\pi \hbar \omega_{\rm c}/\epsilon^\ast V}$.
The momentum dependence of the dipole coupling is neglected here. Instead, a momentum cutoff $k_{\rm c}$ is introduced so as to restrict the electronic states contributing to the polariton formation to $|k|<k_{\rm c}$. It is smaller than or roughly equal to the inverse of lattice spacing (e.g., $ 60/a_0$ for a GaAs-based microcavity, with $a_0$ being the exciton Bohr radius).
%%%%%%%%%%%%%%%%%%%%%%%%%%%%%%%%%%%%%%%%%%%%%%%%%%%%%%%%%%%%%%%%%%%%%%%%%%
%%%%%%%%%%%%%%%%%%  Energy per excitation begin %%%%%%%%%%%%%%%%%%%%%%%%%%
%%%%%%%%%%%%%%%%%%%%%%%%%%%%%%%%%%%%%%%%%%%%%%%%%%%%%%%%%%%%%%%%%%%%%%%%%%
%%%%%%%%%%%%%%%%%%%%%%%%%%%%%%%%%%%%%%%%%%%%%%%%%%%%%%%%%%%%%%%%%%%%%%%%%%
\begin{figure}[tb] 
\begin{center}
\includegraphics[scale=0.39]{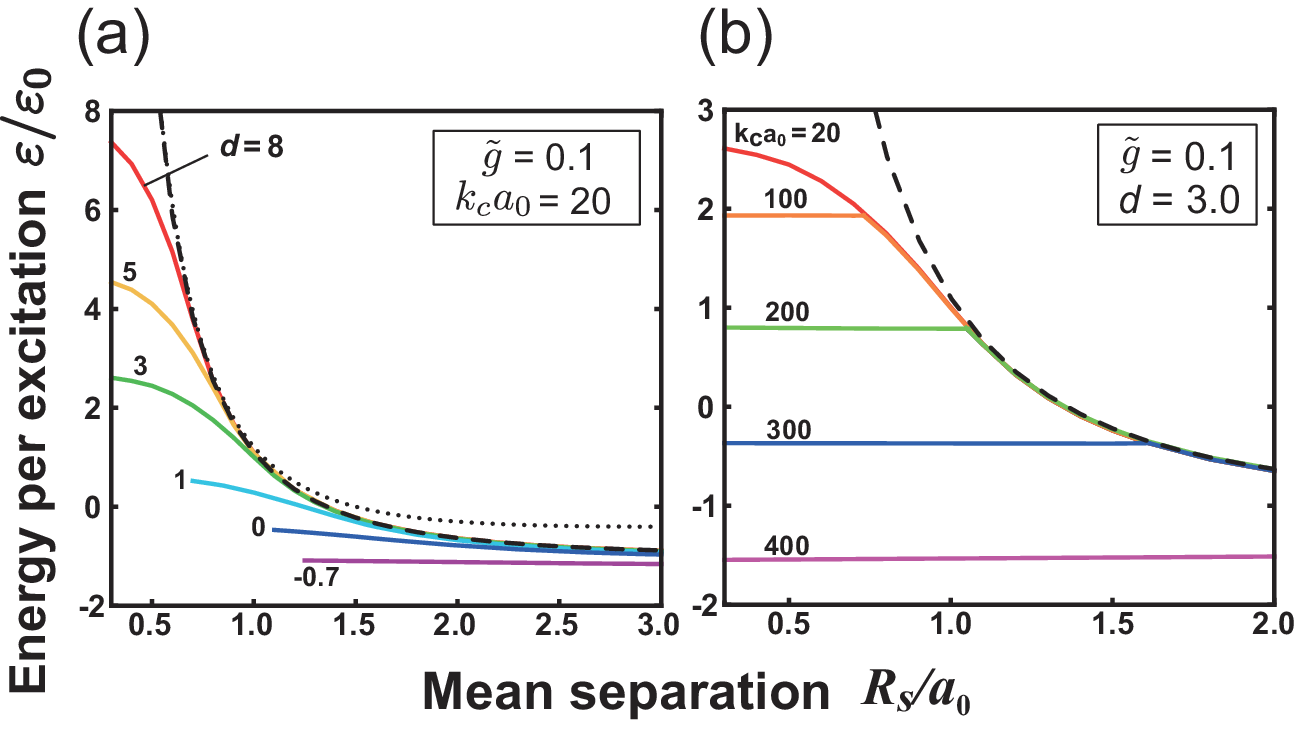} 
\caption{ \label{energy} Energy per excitation is plotted as a function of $R_{\rm s}$: (a) $d =8,5,3,1,0,-0.7$ with $k_{\rm c} a_0=20$, $\tilde{g}=0.1$ (solid) and (b) $k_{\rm c} a_0=20,100,200,300,400$ with $d=3$ and $\tilde{g}=0.1$~(solid). The dashed curves in both figures are obtained for eh systems~\cite{Comte}. The dotted curve in (a) denotes the Hartree-Fock energy for eh plasma.}
\end{center}  
\end{figure}
%%%%%%%%%%%%%%%%%%%%%%%%%%%%%%%%%%%%%%%%%%%%%%%%%%%%%%%%%%%%%%%%%%%%%%%%%%
%%%%%%%%%%%%%%%%%%  Energy per excitation end   %%%%%%%%%%%%%%%%%%%%%%%%%%
%%%%%%%%%%%%%%%%%%%%%%%%%%%%%%%%%%%%%%%%%%%%%%%%%%%%%%%%%%%%%%%%%%%%%%%%%%
%%%%%%%%%%%%%%%%%%%%%%%%%%%%%%%%%%%%%%%%%%%%%%%%%%%%%%%%%%%%%%%%%%%%%%%%%%
Considering the coherent state of polarizations and photons, the mean-field ground state of a polariton condensate is given by
\begin{eqnarray}
|\Phi \rangle&=&e^{(\lambda \psi^\dagger _0-\lambda \psi _0)} \prod_k (e^{i \chi_k} u_k+v_k a^\dagger_k b_k) |{\rm vac} \rangle,  \label{eq:wavefunction}  
\end{eqnarray} 
where $|{\rm vac} \rangle$ denotes the vacuum state with no conduction electrons, no valence holes, and no excited photons.
There is a normalization condition, $u_k^2+v_k^2=1$, that allows us to include the phase-space filling effects of fermions.
The variational parameters $\lambda$, $\chi_k$,  $u_k$, and $v_k$ are determined by the minimization of the total energy $E \left( =\langle H+\mu N_{\rm ex}\rangle \right)$ for a fixed $N_{\rm ex}$, which is given by the expression $\sum_k \langle \psi_k^\dagger \psi_k+(a_k^\dagger a_k +b_k b_k^\dagger)/2 \rangle$.
In the coherent state, all the eh pairs are found to have the same phase: $\chi_k=0$~\cite{Eastham}.
After angular integration, the mean-field energy per excitation $\varepsilon(=E/N_{\rm ex})$ and the total excitation density $\rho_{\rm ex}(=N_{\rm ex}/V)$ are given by
\begin{eqnarray} 
&& \varepsilon / \varepsilon_0
= \frac{R_{\rm s}^3}{a_0^3} \left( 
d \tilde{\lambda}^2+\frac{2}{3 \pi}\int_0^{\kappa_{\rm c}} \!\!\!  \kappa^4 v_k^2 {\rm d}\kappa
-\tilde{g} \tilde{\lambda} \int_0^{\kappa_{\rm c}} \!\!\!  \kappa^2 u_k v_k {\rm d}\kappa 
\right. 
\nonumber  \\
&& \  \left. -   \int_0^{\kappa_{\rm c}} \!\!\! \int_0^{\kappa_{\rm c}} \!\!\! Q_{\kappa_1, \kappa_2} (v_{k1}^2 v_{k2}^2+u_{k1}v_{k1} u_{k2} v_{k2} )
 {\rm d}\kappa_1 {\rm d}\kappa_2 \right) ,   \quad \\
&& \rho_{\rm ex} =\frac{3}{4 \pi a_0^3} \left( \tilde{\lambda}^2 +\frac{2}{3 \pi} \int_0^{\kappa_{\rm c}} \!\!\! \kappa^2 v^2_k~{\rm d} \kappa   \right)=\frac{3}{4 \pi R_{\rm s}^3}, 
\end{eqnarray}
where $\kappa = k a_0$, $\kappa_{\rm c}=k_{\rm c}a_0$, $Q_{\kappa_1, \kappa_2}=\frac{4\kappa_1 \kappa_2}{3 \pi^2}  \ln \left|\frac{\kappa_1+\kappa_2}{\kappa_1-\kappa_2}\right|$, $ R_{\rm s}$ is the mean separation, $\tilde{\lambda}$ is the normalized photon field given by $\tilde{\lambda} = \lambda \sqrt{4\pi a_0^3/3V}$, and $\tilde{g}\left( =g \sqrt{3V/4\pi a_0^3 \varepsilon_0^2} \right)$ is a dimensionless coupling constant.
It is difficult to determine an infinite number of variational parameters, we apply for the excitonic constituent the interpolating wave function proposed by Comte and Nozieres~\cite{Comte}:
\begin{eqnarray}
\frac{u_k v_k}{u_k^2-v_k^2}=\frac{\zeta}{(1+\kappa^2)(\kappa^2-\Omega)}. \label{eq:comte3d}
\end{eqnarray}
Our task is to determine three parameters $\zeta$, $\Omega$, and $\lambda$.

%%%%%%%%%%%%%%%%%%%%%%%%%%%%%%%%%%%%%%%%%%%%%%%%%%%%%%%%%%%%%%%%%%%%%%%%%%
%%%%%%%%%%%%%%%%%%  Photon field begin %%%%%%%%%%%%%%%%%%%%%%%%%%%%%%%%%%%
%%%%%%%%%%%%%%%%%%%%%%%%%%%%%%%%%%%%%%%%%%%%%%%%%%%%%%%%%%%%%%%%%%%%%%%%%%
%%%%%%%%%%%%%%%%%%%%%%%%%%%%%%%%%%%%%%%%%%%%%%%%%%%%%%%%%%%%%%%%%%%%%%%%%%
\begin{figure}[tb]
\begin{center} 
\includegraphics[scale=0.38]{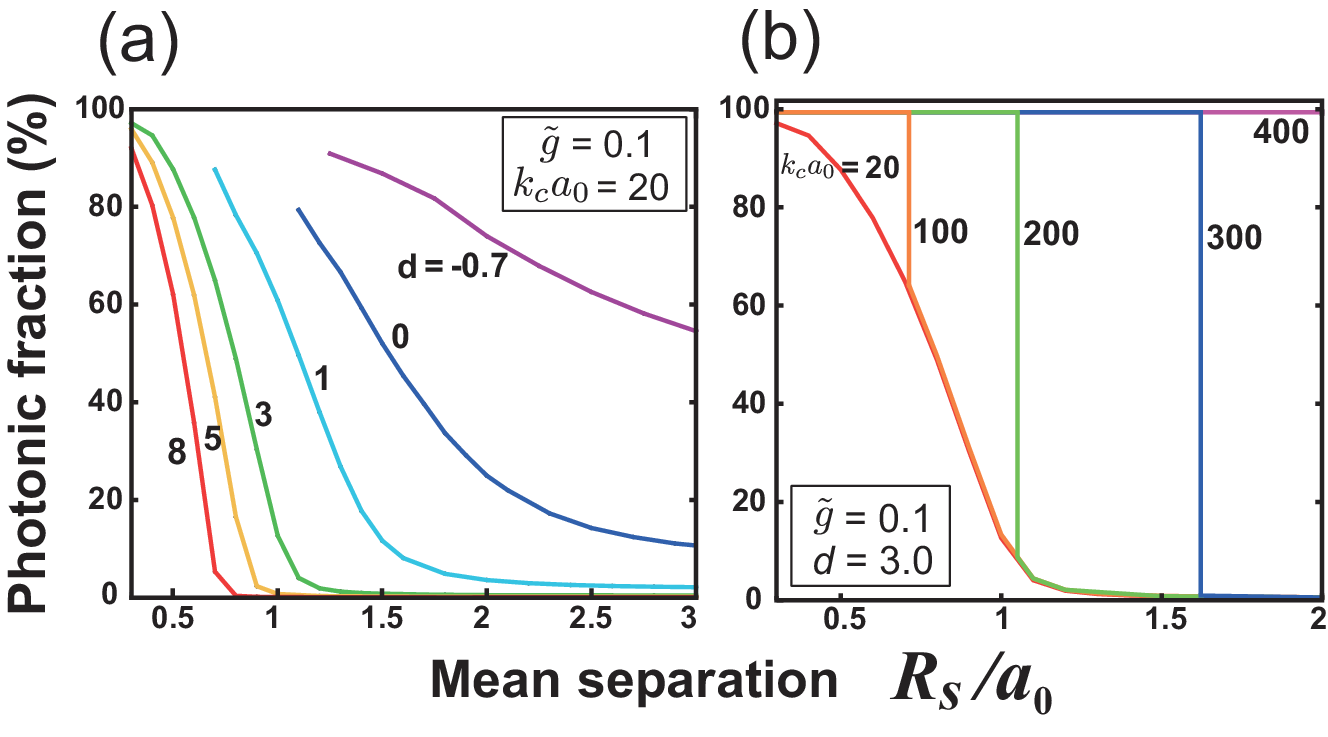} 
\caption{ \label{photon} Photonic fraction plotted as a function of $R_{\rm s}$ for the same parameter sets as in Figs.~\ref{energy}(a) and \ref{energy}(b), respectively.}
\end{center} 
\end{figure}
%%%%%%%%%%%%%%%%%%%%%%%%%%%%%%%%%%%%%%%%%%%%%%%%%%%%%%%%%%%%%%%%%%%%%%%%%%
%%%%%%%%%%%%%%%%%%  Photon field end   %%%%%%%%%%%%%%%%%%%%%%%%%%%%%%%%%%%
%%%%%%%%%%%%%%%%%%%%%%%%%%%%%%%%%%%%%%%%%%%%%%%%%%%%%%%%%%%%%%%%%%%%%%%%%%
Figure~\ref{energy} shows the mean-field energy per excitation plotted as a function of $R_{\rm s}$.
In Fig.~\ref{energy}(a), each curve (colored lines) corresponds to a different value of $d$; however, all the curves have the same cutoff parameter $k_{\rm c} a_0=20$.
The curves approach that of an eh system (dashed) in the low-density region, except near resonance, where $d \sim -1$ $(\hbar \omega_{\rm c} \sim E_{g}-\varepsilon_{0})$, and in the high-density region, slightly below the photon level $d$. The energy shift from $\varepsilon/\varepsilon_0 = -1$ to $\varepsilon/\varepsilon_0 =d$ indicates the polariton condensation crossovers from excitonic to photonic ones with an increase in the density.
The energy saturation at high density is explained by fermionic phase-space filling. The number of eh pairs increases until the conduction electron band is filled up to the photon level, and thereafter, photonic excitations replace those of eh pairs to minimize the total energy.
The photonic character is observed above a density ($R_{\rm s}<R_{\rm s}^\ast$) where the curves move away from the dashed curve.
The crossover can be seen directly from the plots in Fig.~\ref{photon}. Figure~\ref{photon}(a) is obtained for the same parameter set as Fig.~\ref{energy}(a).
The photonic fraction increases sharply from zero for $R_{\rm s} < R_{\rm s}^\ast$, except for $d \sim -1$, where a large photonic fraction is present at low densities.
Slopes at $ R_{\rm s} < R_{\rm s}^\ast$ are higher for larger values of $d$. The evolution of the state from excitonic to photonic is always smooth for the cutoff value of $k_{\rm c} a_0=20$.

In the excitonic regime where $R_{\rm s}>R_{\rm s}^\ast$, the ground state is classified into three kinds even though there are no clear boundaries. If one-excitation energy of eh systems (dashed in Figs.~\ref{energy}(a) and (b)) is close to the sigle-exciton level $-\varepsilon_0$, the eh pairs can be regarded as the BEC of excitons. The low-density regime $(R_{\rm s} \gtrsim 2)$ is categorized as ``exciton BEC." The highest density regime $(R_{\rm s} \lesssim 1)$ is categorized as ``eh plasma"~\cite{note1} since the curve of the eh system (dashed curve in Figs.~\ref{energy}(a) and (b)) overlaps with that of eh plasma (dotted curve in Fig.~\ref{energy}(a)), where eh pairs are unbound and all excitations are fermionic. For the intermediate-density $(1 \lesssim R_{\rm s} \lesssim 2)$, which is categorized as ``eh BCS," eh pairs are regarded as weakly bound fermions like Cooper pairs of BCS superconductivity.
Since $R_{\rm s}^\ast$ varies, $d$ determines the regimes the polariton system passes through---exciton BEC, eh BCS, and eh plasma---before the ground state of the system becomes photon-like.

The plots shown in Figs.~\ref{energy}(a) and \ref{photon}(a) are shown in Figs.~\ref{energy}(b) and \ref{photon}(b) for a different paramter set i.e., for $d=3$ and several values of $k_{\rm c}$. 
The change in the polariton state at $k_{\rm c} a_0=100,200,300$ is not a crossover but rather a first-order transition since there is a discontinuous jump in the slope of the curve in Fig.~\ref{energy}(b). Similarly, as shown in Fig.~\ref{photon}(b), the photonic fraction jumps to almost 100$\%$ at the transition density. The system ground state is described by another type of solution different from the one that approaches the ground state of the eh system at low densities. We shall call the former {\it photon solution} since the photonic fraction is almost 100$\%$, and the latter {\it polariton solution}.
For $k_{\rm c} a_0=400$, the photon solution corresponds to the ground state for all densities. 

\begin{figure}[tb]
\begin{center}
\includegraphics[scale=0.7]{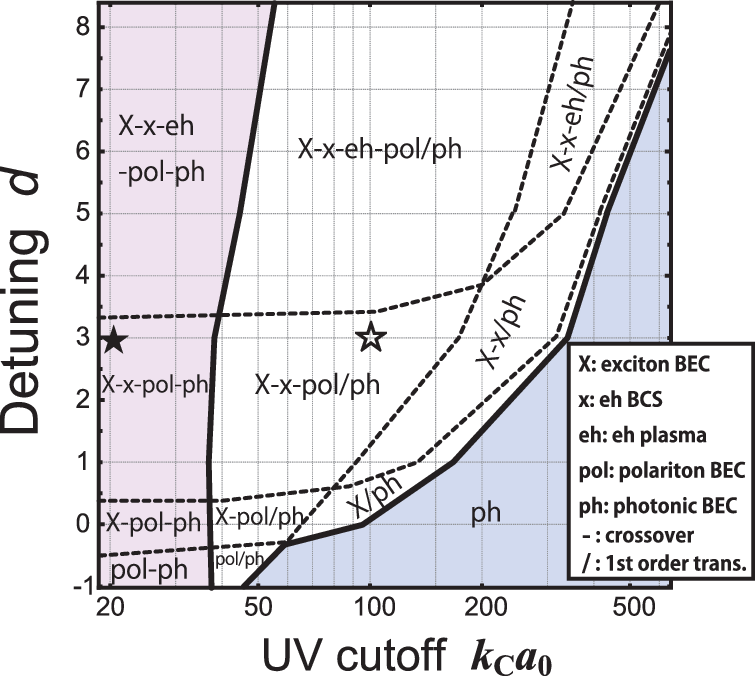}
\caption{ \label{map} Phase diagram showing the crossover properties of the polariton BEC when the excitation density increases. The coupling parameter is set as $\tilde{g}=0.1$. Phase boundaries that are clear and unclear are shown by the solid and the dashed lines, respectively (see the text for an explanation).}
\end{center}
\end{figure}

As discussed above, the evolution of the ground state, with increasing excitations, depends on $d$ and $k_{\rm c}$.
Various types of the evolution are summarized in the phase diagram of Fig.~\ref{map}.
The labels X, x, eh, and pol denote the different regimes where the polariton solution corresponds to the ground state: exciton BEC, eh BCS, eh plasma, and polariton BEC, respectively~\cite{cf1}.
The label ph stands for the regime where the ground states are photon-like.
Therefore, for the parameter regime marked as ``X-x-pol/ph'' at the center of Fig.~\ref{map}, the condensation changes from/to exciton BEC, eh BCS, polariton BEC, and photonic BEC in increasing order of excitations. The hyphen (-) and slash (/) indicate that the change between the different regimes is a crossover and a first-order transition, respectively. Boundaries between different phases are determined under certain conditions~\cite{cf1}. The boundaries that depend on the conditions are shown as dashed lines.
The diagram indicates the following: (i) The larger the value of $k_{\rm c}$, the stronger the photonic nature of the condensate. (ii) The lesser the value of $k_{\rm c}$ and $d$, the stronger the coupling between the eh system and photons, i.e., the more polaritonic is the condensate.
%%%%%%%%%%%%%%%%%%%%%%%%%%%%%%%%%%%%%%%%%%%%%%%%%%%%%%%%%%%%%%%%%%%%%%%%%%
%%%%%%%%%%%%%%%%%%  Phase diagram begin %%%%%%%%%%%%%%%%%%%%%%%%%%%%%%%%%%%%%%%
%%%%%%%%%%%%%%%%%%%%%%%%%%%%%%%%%%%%%%%%%%%%%%%%%%%%%%%%%%%%%%%%%%%%%%%%%%
%%%%%%%%%%%%%%%%%%%%%%%%%%%%%%%%%%%%%%%%%%%%%%%%%%%%%%%%%%%%%%%%%%%%%%%%%%
%%%%%%%%%%%%%%%%%%%%%%%%%%%%%%%%%%%%%%%%%%%%%%%%%%%%%%%%%%%%%%%%%%%%%%%%%%
%%%%%%%%%%%%%%%%%%  Phase diagram end   %%%%%%%%%%%%%%%%%%%%%%%%%%%%%%%%%%%%%%%
%%%%%%%%%%%%%%%%%%%%%%%%%%%%%%%%%%%%%%%%%%%%%%%%%%%%%%%%%%%%%%%%%%%%%%%%%%

\begin{figure}[tb]
\begin{center}
\includegraphics[scale=0.75]{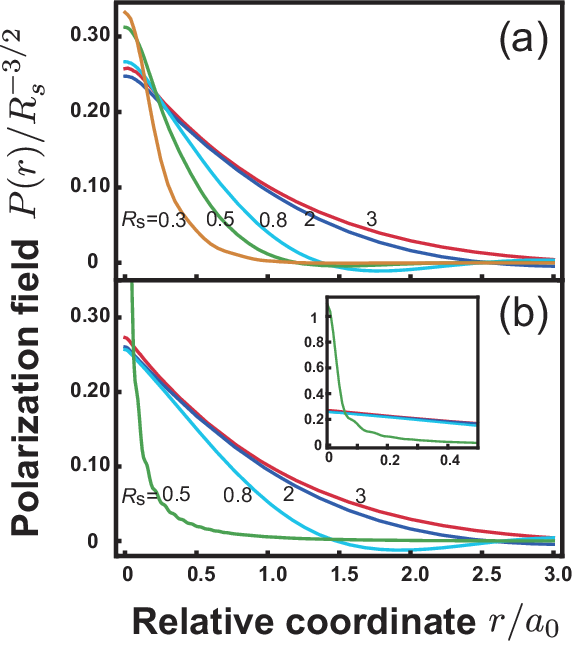}
\caption{ \label{polarization} Real space profiles of the eh wave function $P(r)/R_{\rm s}^{-3/2}$ for (a) $R_{\rm s}=3, 2, 0.8, 0.5, 0.3$ with $k_{\rm c} a_0 =20$ and (b) $R_{\rm s}=3, 2, 0.8, 0.5$ with $k_{\rm c} a_0 =100$. The inset shows a magnified view of the region $r/a_0<0.5$. We set $\tilde{g}=0.1$ and $d=3$ for both graphs.}
\end{center}
\end{figure}

To scrutinize the difference between the polariton solution and the photon solution, we plot in Fig.~\ref{polarization} the wave function of an eh pair, $P(r)=(1/V) \sum_k \langle a_k^\dagger b_k \rangle \exp({\rm i}kr)$, as a function of the relative coordinate $r$ between an electron and a hole.
Plots are obtained for various densities ($R_{\rm s}$) for $d=3$ and $k_{\rm c}a_0=20$ [Fig.~\ref{polarization}(a)] and for $d=3$ and $k_{\rm c} a_0=100$ [Fig.~\ref{polarization}(b)].
In the former figure, the wave function changes its shape and narrows gradually with increasing density from the wave funtion of 1s exciton, indicating that the parameter used (š in Fig.~\ref{map}) is in the regime where all the expected changes in the ground state are crossovers.
While the width $a_0$ is determined by the Coulomb attraction in the exciton BEC regime, it is modified by the electric dipole interaction for $R_{\rm s}<R_{\rm s}^\ast$, where the photonic character appears.
The change in the binding force is seen more clearly in Fig.~\ref{polarization}(b)~(™ in Fig.~\ref{map}).
The wave function gradually narrows as $R_{\rm s}$ decreases from 3 to 0.8, and it shows a drastic change to have a sharp peak at $r=0$ for $R_{\rm s} \lesssim 0.7$, i.e., in the photonic BEC regime. Clearly, the width is determined by a new length scale and not by $a_0$, which indicates that the mechanism of eh pairing is completely different from that in the case of dilute excitons.
We find that the photon solution satisfies the conditions $v_k \ll 1 $, $u_k \to 1$, and $\tilde{\lambda} \neq 0$. Therefore, the variational equation for the eh wave function $P(r)$ reduces to
\begin{eqnarray} 
\left(-\frac{\hbar^2 \nabla^2}{2m_{r}} -\frac{4\pi e^2}{\epsilon^\ast}\frac{1}{|r|}
 \right)P(r)-g \lambda \tilde{\delta}(r)=\mu P(r), \label{eq:polarizaion}
\end{eqnarray} 
where $\tilde{\delta}(r) \left[ = (1/V)\sum_{|k|<k_{\rm c}}\exp(ikr) \right]$ is localized at $r=0$ with a width of $\sim 1/k_{\rm c}$ and becomes a delta function if $k_{\rm c} = \infty$.
The reduced mass of an eh pair is denoted by $m_{r}$.
Since the third term on the left side can be rewritten as $-\frac{g^2 V}{2(d-\mu)}\tilde{\delta} (r) P(r)$, we see that the photon field induces an attractive delta potential mediated by photons~\cite{Marchetti1}.
When considering two or three dimensions, the lowest bound-state energy of the real delta potential becomes minus infinity.
However, a momentum cutoff was introduced here, hence the bound-state energy remains finite. The energy is estimated as $\varepsilon/\varepsilon_{0} \sim d - \frac{3 \pi}{8} \tilde{g}^2 k_{\rm c} a_0$, which is obtained using some approximations and the conditions $v_k \ll 1$, $\Omega \to -\infty$, and $R_{\rm s}^{3/2} |\zeta/\Omega| \lesssim 1 $~(the photon solution satisfies them).
We can conclude that the first-order transition occurs when the bound-state energy due to the short-range attraction falls below the energy of the polariton solution.
The solution of Eq.~(\ref{eq:polarizaion}) for large $k_{\rm c}$ is
% given by 
\begin{eqnarray}
P(r)=\tilde{g}\tilde{\lambda} 
\left( \sum_{n=1}^{\infty} \frac{\varphi_{n{\rm s}}^{{\rm b} \ast}(0)\varphi_{n{\rm s}}^{{\rm b}}(r)}{(E_{n{\rm s}}^{{\rm b}}-\mu)/\varepsilon_0}
+\sum_{|k|<k_{\rm c}} \frac{\varphi_{k}^{{\rm s}\ast}(0)\varphi_{k}^{{\rm s}}(r)}{(E_{k}^{{\rm s}}-\mu)/\varepsilon_0}
\right), \quad 
\end{eqnarray}
where $\varphi_{n{\rm s}}^{{\rm b}}$ and $E_{n{\rm s}}^{\rm b}$ ($\varphi_{k}^{{\rm s}}$ and $E_{k}^{\rm s}$) are the wave functions and the energy of $n$s-exciton bound state (scattering state with momentum $k$). 
As discussed above, the chemical potential has an upper limit, i.e., $\mu /\varepsilon < d$.
For a large momentum, the main contribution to $P(r)$ comes from the scattering states with $P(k) \propto k^{-2}$, and $P(k)$ is long-tailed in the photonic regime compared to the case of the dilute limit of 1s excitons $[P(k)\propto k^{-4}]$. This directly leads to the narrowing of $P(r)$.
%%%%%%%%%%%%%%%%%%%%%%%%%%%%%%%%%%%%%%%%%%%%%%%%%%%%%%%%%%%%%%%%%%%%%%%%%%
%%%%%%%%%%%%%%%%%%%%%%%%%%%%%%%%%%%%%%%%%%%%%%%%%%%%%%%%%%%%%%%%%%%%%%%%%%
%%%%%%%%%%%%%%%%%%  Polarization field begin %%%%%%%%%%%%%%%%%%%%%%%%%%%%%
%%%%%%%%%%%%%%%%%%%%%%%%%%%%%%%%%%%%%%%%%%%%%%%%%%%%%%%%%%%%%%%%%%%%%%%%%%
%%%%%%%%%%%%%%%%%%%%%%%%%%%%%%%%%%%%%%%%%%%%%%%%%%%%%%%%%%%%%%%%%%%%%%%%%%
%%%%%%%%%%%%%%%%%%%%%%%%%%%%%%%%%%%%%%%%%%%%%%%%%%%%%%%%%%%%%%%%%%%%%%%%%%
%%%%%%%%%%%%%%%%%%  Polarization field end   %%%%%%%%%%%%%%%%%%%%%%%%%%%%%
%%%%%%%%%%%%%%%%%%%%%%%%%%%%%%%%%%%%%%%%%%%%%%%%%%%%%%%%%%%%%%%%%%%%%%%%%%
%%%%%%%%%%%%%%%%%%%%%%%%%%%%%%%%%%%%%%%%%%%%%%%%%%%%%%%%%%%%%%%%%%%%%%%%%%

Our result can give clues to build theretical models that can appropriately describe the BEC.
Since the momentum cutoff introduced here will be of the same order as the inverse of the lattice constant, the width of the eh wave function can be of the same order as the lattice constant in the photonic regime.
In such a case, an eh pair should be recognized as a Frenkel exciton, and hence, the description is beyond the capability of our model which employs the effective mass approximation.
This indicates the need for other theoretical models to treat localized excitons such as the Dicke model~\cite{Eastham,Keeling}.

In summary, the mean-field ground state of a microcavity polariton system is determined by a variational approach~\cite{Comte}.
The ground state changes from excitonic to photonic regime.
In the photonic regime, eh pairs are shown to have a small radius since they are bound by photon-mediated delta attraction.
With increasing excitations, various types of ground-state evolutions are expected depending on the detuning and momentum cutoff.
The phase diagram in Fig.~\ref{map} summarizes the evolutions; the change is a crossover when the cutoff momentum is not too large, and the first-order transition from excitonic to photonic condensation is expected when the cutoff is large.
Although the calculation presented here is for three dimensions, the same scenario applies to two dimensions, which will be shown elsewhere.

The authors are grateful to Kenichi Asano, Takuma Ohashi, and Kouta Watanabe for fruitful discussions. This work was supported by KAKENHI (20104008).


\begin{thebibliography}{99}
\bibitem{note1} Strictly speaking, this ``eh plasma'' may differ from a normal state at a finite temperature since the pair gap remains finite however small it might be.
\bibitem{cf1} 
$R_{\rm s}^\ast$ is defined by the density above which the photonic fraction becomes larger than 20$\%$ in case of the polariton solution.
At low density ($R_{\rm s}>R_{\rm s}^\ast$), boundaries between exciton BEC, eh BCS, and eh plasma are given by $R_{\rm s}=2$ and $R_{\rm s}=1$.
We defined ``polariton BEC,'' which is marked as ``pol'' in Fig.~\ref{map}, as the density region where the photonic fraction is larger than 20$\%$ and less than 80$\%$. If the fraction is larger than 80$\%$, the BEC is regarded to be photon-like, and the corresponding regime is marked as ``ph'' in Fig.~\ref{map}.
\end{thebibliography}
\end{document}